\documentstyle[preprint,epsfig,aps]{revtex}

\begin{document}

\centerline{\bf LIMITS ON TOPOLOGICAL DEFECT NEUTRINO FLUXES} 

\centerline{\bf FROM HORIZONTAL AIR SHOWER MEASUREMENTS}

\centerline{ J.J.~Blanco-Pillado, R.A.~V\'azquez$^*$ and E.~Zas }

\centerline{\it Departamento de F\'\i sica de Part\'\i culas, Universidad de
Santiago}
\centerline{\it E-15706 Santiago de Compostela, Spain}
\centerline{$^*$\it INFN sez. di Roma and Universit\'a della Basilicata, Italy}

\begin{abstract}

We obtain the horizontal air shower rate from the (EeV) high energy neutrino
flux predicted in some topological defect scenarios as the source for the
highest energy cosmic rays. Emphasis is made on the different character of the
events depending on the neutrino flavor an type of interaction. We show that
the bound for muon poor showers in the $10^5-10^7$ GeV energy range is violated
by maximal predictions for superconducting cosmic string neutrino fluxes, we
compare it to other neutrino flux limits and we discuss the future of such
measurements to further constrain these models.

\end{abstract}

The establishment of cosmic rays of energies above $10^{20}~$eV\cite{Bird}  and
the prospects for the construction of a giant air shower array for their study,
the Pierre Auger project\cite{Auger}, has stimulated great activity in the
search of theoretical models for the production of these particles. The
annihilation of topological defects (TD's) has been recently proposed as the
origin of the highest energy cosmic rays\cite{Witten} although there is some
controversy about this possibility  \cite{Kibble_1}. This mechanism avoids the
main difficulties in the conventional shock acceleration and would imply large
fluxes of gamma rays and neutrinos up to energies in the EeV range and well
above\cite{Sigl_1}. Active Galactic Nuclei (AGN), the most powerful known
objects, have also been suggested as the possible origin of the highest energy
cosmic rays\cite{AGNsource}. If the very high energy gamma rays detected from
AGN\cite{EGRET} come from pion decay, in proton acceleration models, they must
also emit high energy neutrinos. Models for acceleration in the AGN jets 
predict neutrinos also extending to the EeV range\cite{Manh95,Manh96,Proth96}. 

Some of these models have already been shown to predict secondary photon and
neutrino fluxes in conflict with experiment. Gamma ray searches ($\sim
100$~MeV) constrain certain combinations of extragalactic magnetic fields and
TD models because of electron positron pair production in the magnetic
field\cite{Protheroe_2}. Also the Fr\`ejus underground muon detector has not
observed the muon neutrino prediction for  Superconducting Cosmic Strings
(SCS)\cite{Frejus}. Clearly there are many uncertainties in these models such
as the normalization, the mass of the X particle in the GUT scale, M$_X$, the
extragalactic magnetic field, the fragmentation functions involved as well as
the propagation of each of the  produced particles in the extragalactic
magnetic field and background  radiations, taking into account the evolution of
the Universe. It has been claimed by different groups \cite{Protheroe,Coppi}
that tuning the free parameters in the model (M$_X$, the maximum of the
injection spectrum and the extragalactic magnetic field), it is still possible
to explain the UHE cosmic events, avoiding the 100 MeV gamma ray constraints
\cite{Chi_1}. 

Large scale high energy neutrino detectors are expected to further test these
models\cite{AGNsource}. As the earth becomes opaque for neutrinos already in
the 100 TeV to 1 PeV range, EeV neutrinos can only be detected for zenith
angles between vertical down going and horizontal. In such case the experiment
must have some control over the energy of the muon to separate the neutrino
signal from that of atmospheric muons which poses a serious background.
Underground muon experiments rely on their overburden for this purpose. For
high energies they must search for close to horizontal neutrinos where the
overburden and energy uncertainties are largest. Detectors in water or ice may, 
in addition, detect \v Cerenkov light from neutrino showers with the advantage
of being also sensitive to the electron neutrino. They require good directional
and shower reconstruction capabilities to separate neutrino showers from those
induced by muons. Horizontal Air Shower (HAS) measurements provide a third
alternative as they are due to high energy penetrating particles such as muons
and neutrinos\cite{Berezinsky}, including electron neutrinos, and shower size
determinations can be used as an energy threshold. Existing air shower arrays
can  constrain neutrino fluxes\cite{Halzen_2,Baltrusaitis} and their relevance
for direct neutrino detection will become more important when the Pierre Auger
project is constructed\cite{ParenteZas,Cronin}.

In this paper we calculate the HAS rate due to neutrino fluxes from TD and
recent AGN models separating explicitly the muon poor HAS in order to compare
it to the bound published by the AKENO group\cite{AGASA}. The $\nu _e+ \bar
\nu_e$ flux prediction in the maximal superconducting string model violates 
this bound. We compare the different shower channels in this model showing that
about half the total  $\nu_e+ \bar \nu_e$ contribution is muon poor. The muons
produced by charged current muon neutrino interactions in the atmosphere
dominate the ''standard'' atmospheric muon flux at high energies and we
calculate the HAS contribution from secondary bremsstrahlung from these muons.
This is the dominant contribution to muon poor showers from $\nu_{\mu}+ \bar
\nu_{\mu}$. We finally compare the AKENO bound to other bounds emphasizing the
differences in energy range and we briefly address the potential capabilities
of typical existing and planned air shower arrays to detect neutrino induced
horizontal showers.

{\sl Horizontal Air Showers.}
Extensive air showers initiated by primary cosmic rays are strongly suppressed
at large zenith angles because the slant depth of the atmosphere rises from
$\sim$ 30 radiation lengths in the vertical direction to $\sim$ 1,000 in the
horizontal direction and the absorption depth of showers only rises
logarithmically with primary energy. It has been known for long that only
weakly interacting particles, such as muons and neutrinos, may penetrate such
large depths to interact and produce a shower sufficiently close to be detected
at ground level. HAS have been observed in the 60's by several groups for
shower sizes between 10$^{3}$-10$^{5}$ particles \cite{Tokyo}. The shower rate
observed is consistent with conventional muon production by hard bremsstrahlung
\cite{Tokyo,Kiraly,Halzen_2,Parente}. For shower sizes above 10$^{5}$
particles, AKENO has published an upper bound on muon poor air showers at
zenith angles greater than 60$^{0}$ \cite{AGASA}. These results have been used
as a bound for the high energy muon fluxes induced by the semileptonic decay of
charm particles produced by primary cosmic rays in the atmosphere
\cite{Halzen_2}, and as a non-accelerator hint on the very high energy charm
cross section \cite{Gonzalez}.

There are multiple shower channels for neutrino showers. Deep Inelastic
Scattering (DIS) interactions of the neutrino with air nucleons via charged and
neutral currents produce a hadronic shower of energy $yE_{\nu}$, where $y$ is
the energy fraction transferred to the nucleon. In charged current
electron-neutrino (antineutrino) interactions the emitted electron (positron)
generates an electromagnetic cascade carrying the remaining energy of the
initial neutrino $(1-y)E_{\nu}$ that is superimposed to the cascade initiated
by the hadron debris. For muon neutrinos the charged current interaction
produces a secondary flux of high energy muons which could in turn induce
horizontal showers further into the atmosphere in the same way the atmospheric
muons do. The interaction of neutrinos with atomic electrons is in general
suppressed by the ratio of the electron to the proton rest mass except for the
Glashow resonance, $\overline {\nu}_e + e^- \rightarrow W^-$, which dominates
over all other processes at the resonance energy $E_{\bar \nu_e} \sim 6.4 \;
10^6$ GeV\cite{Reno}. The decay of the $W$ boson into an electron and an
antineutrino generates a purely electromagnetic shower with energy $E \sim 3 \;
10^6$ GeV. Its decay in the muon channel will also contribute to the horizontal
muon flux at these energies, while the tau channel will in turn decay into
hadrons $64\%$ of the times and into a muon or an electron $18\%$ of the times
each.

Following ref.\cite{Halzen_2} the differential rate of horizontal cascades at
fixed shower size, $N_{e}$, is obtained similarly for neutrinos and muons
integrating the parent flux, the $y$ differential cross section of the
corresponding interaction and the atmospheric depth:
\begin{equation}
\phi_{sh}(N_e) = \int_0^{\infty} \!\! dt \int_0^1 {dy \over \bar y}
\phi_{\nu,\mu}\left(E_{\nu,\mu}={E(N_e,t) \over \bar y}\right)
{d \sigma \over dy}
{dE \over dN_e} \;,
\label{eq:shower}
\end{equation}
Here $\bar y$ is $y$, $1-y$ or 1 depending on the interaction that produces the
shower. For muon bremsstrahlung  the energy of the photon which originates a
shower at a given depth and of a given shower size is determined, on average,
by the inverse of the Greisen parametrization $E(N_e,t)$ \cite{Greisen}. The
energy of the parent muon is given by $E/y$. The factor $dE/dN_e$ is just the
Jacobian of the transformation. For neutral current interactions the energy of
the hadron debris is inferred inverting the parametrization of hadronic showers
due to Gaisser\cite{Gaisser}. In the case of charged current electron neutrino
interactions, eq.~(\ref{eq:shower}) applies with $\bar y = 1$, because all the
energy is transferred to the shower; and a more complicated shower size
depending on $y$ and combining both parametrizations has been used.

\begin{figure}
\epsfxsize=10cm
\begin{center}
\mbox{\epsfig{file=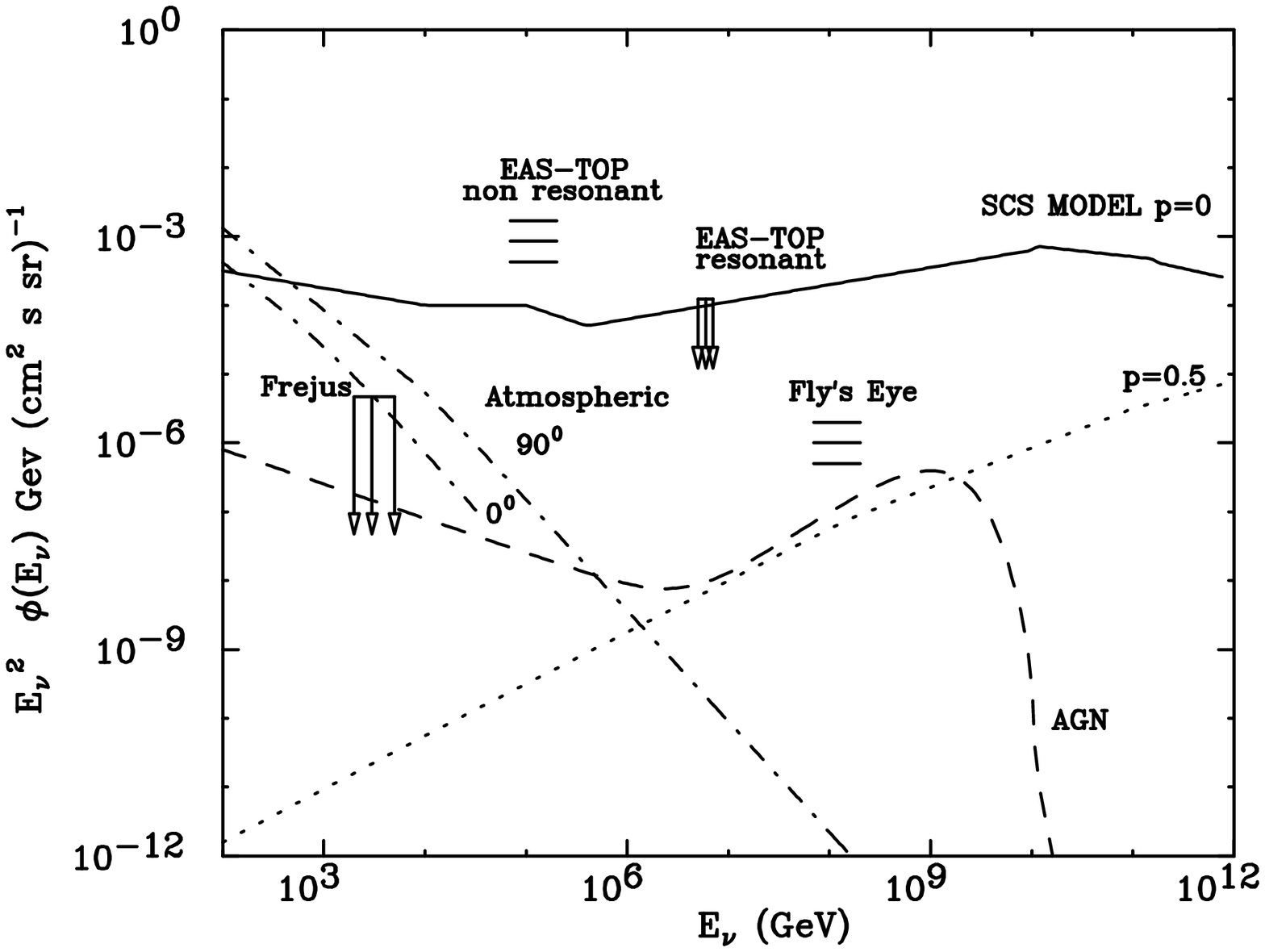}}
\end{center}
\caption{Differential ($\nu_\mu + \bar \nu_\mu$) fluxes of ref.~29  
for $p=0$ (solid line) and for $p=0.5$ (dotted). The dashed line 
is the prediction of ref.~9 
and the dot--dashed line is 
the atmospheric flux. Also shown are some published limits 
as marked. The three parallel lines illustrate the change in the limits for 
different spectral indices (1.1, 2 and 3 from bottom to top).}
\label{fluxes}
\end{figure}

\begin{figure}
\epsfxsize=10cm
\begin{center}
\mbox{\epsfig{file=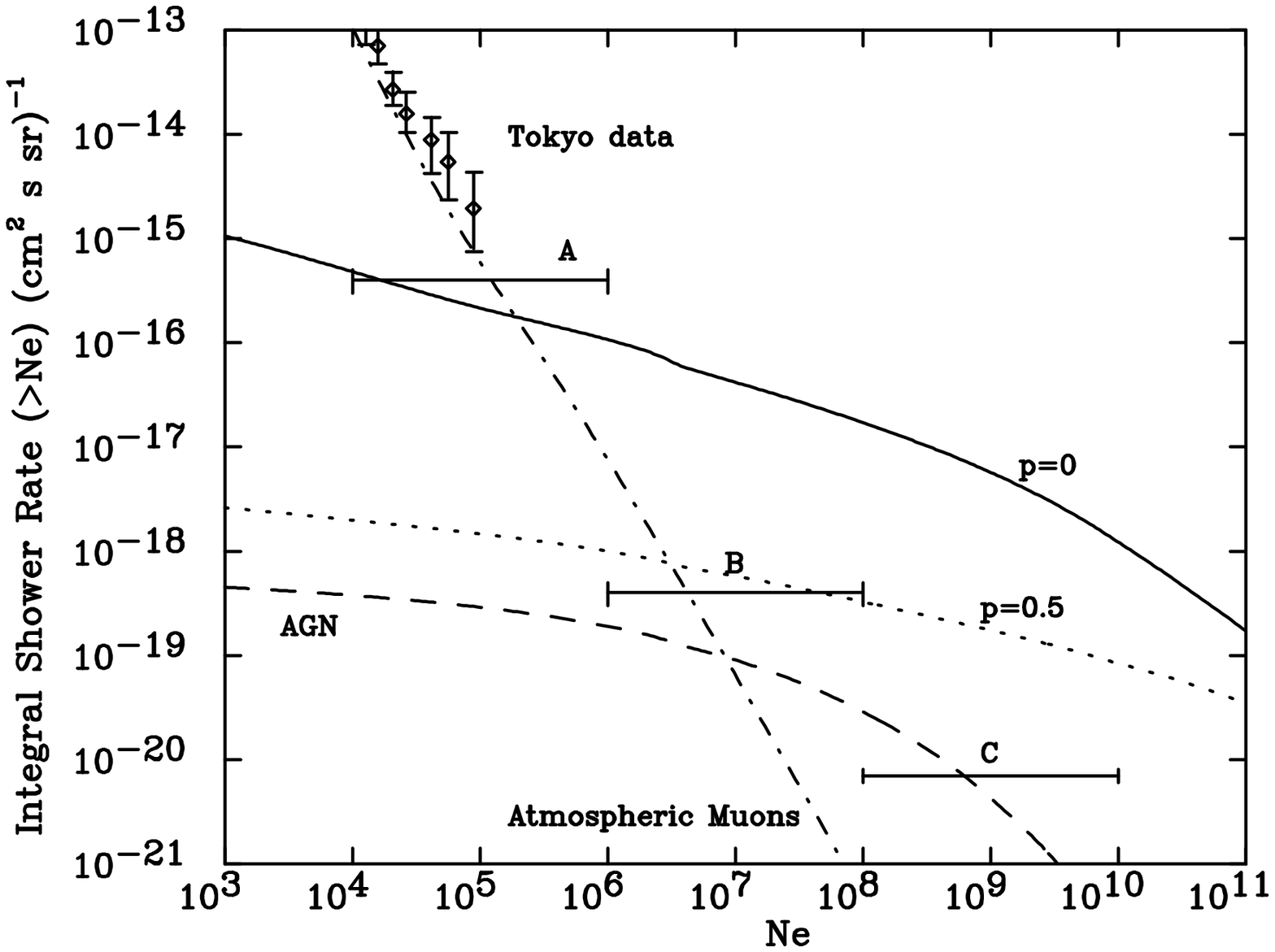}}
\end{center}
\caption{Integral horizontal shower rate as a function of shower  size (Ne) for
different models.  TD neutrinos with $p=0$ (solid line), $p=0.5$ (dotted), AGN
neutrinos (dashed) and  atmospheric muons (dot-dashed). Diamonds represent the
HAS  data from the University of Tokyo. Horizontal lines marked as A, B and C 
represent the expected limit of non observation in a year made with ideal 
detectors of areas 10$^5~m^2$ (A), 10$^8~m^2$ (B), and 10$^{10}~m^2$ (C). See 
text.}
\label{showers}
\end{figure}

We have performed the previous calculations using the neutrino flux 
predictions from topological defects \cite{Bhattacharjee} and from AGN models 
\cite{Manh96}, and we have assumed that the fluxes for $\nu_e$,
$\overline{\nu}_e$, $\nu_{\mu}$ and $\overline{\nu}_{\mu}$ are in the ratio
1:1:2:2. In fig.(\ref{fluxes}) we show the $\nu_{\mu}+ \bar \nu_{\mu}$ 
differential fluxes in these models compared to atmospheric neutrinos. For the
neutrino cross section we have used the MRS(G) parton distribution functions
\cite{Martin} extrapolated to low $x$ as described in \cite{ParenteZas}. We
have checked that our results are not sensitive to using alternative
parametrizations for shower size such as that of Fenyves et al. \cite{Fenyves},
nor to the multiplicity of hadrons in the DIS interactions. In
fig.(\ref{showers}) we show the total integral horizontal air shower rates for
the TD and AGN models as a function of the shower size, compared to Tokyo data
and to expectations from atmospheric muons.

\begin{figure}
\epsfxsize=10cm
\begin{center}
\mbox{\epsfig{file=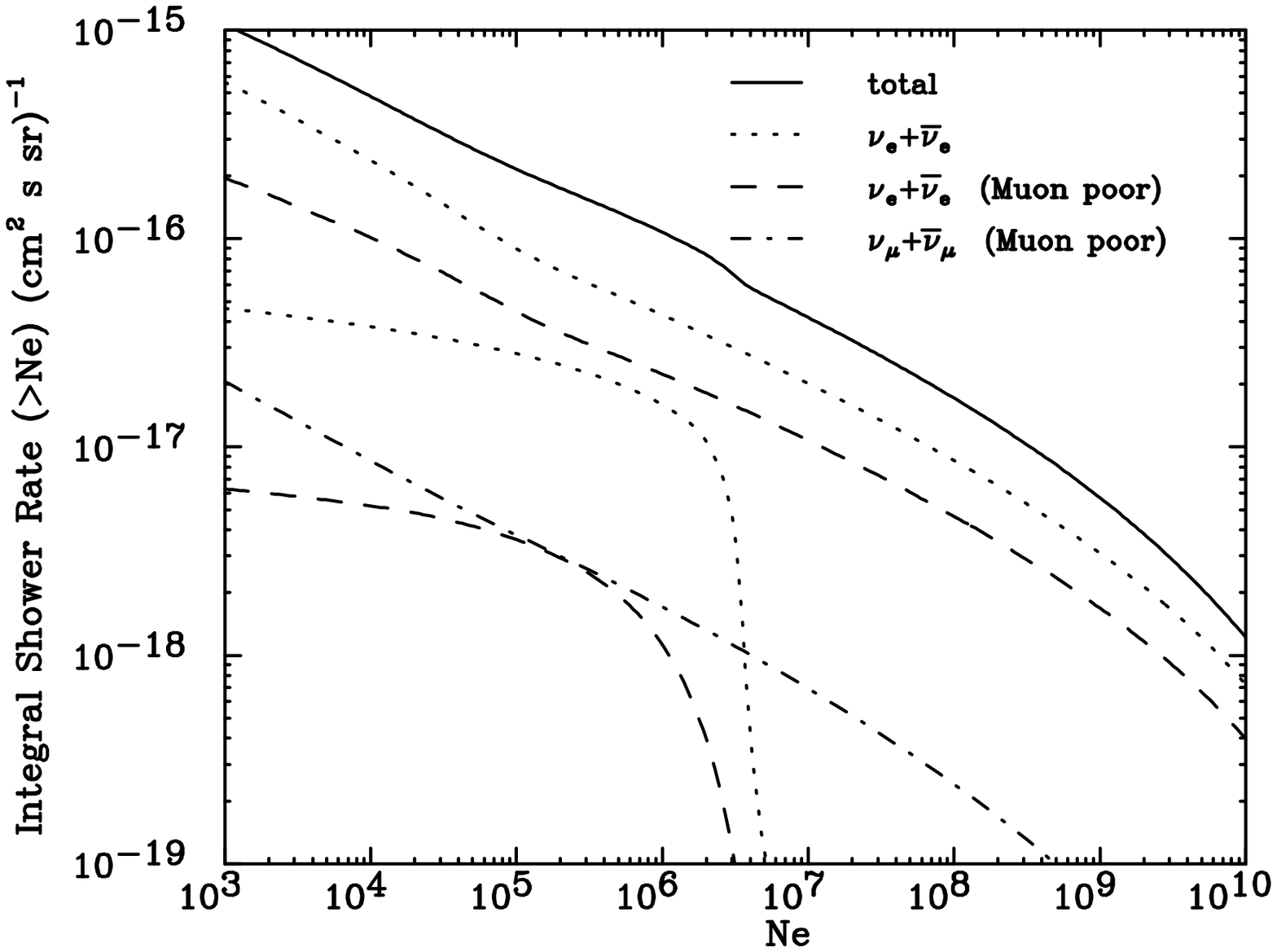}}
\end{center}
\caption{Partial HAS channels for the TD model with $p=0$.  From top to bottom:
total rate, ($\nu_e+\bar \nu_e$) DIS contribution, the muon  poor ($\nu_e+\bar
\nu_e$) DIS contribution, the total resonant contribution  ($\bar \nu_e$), the
secondary muon bremsstrahlung and the resonant contribution  into the electron
channel.} 
\label{contributions}
\end{figure}

{\sl Muon poor showers.}
The AKENO bound on horizontal air showers has been obtained on the basis of
searching for muon poor showers. This has allowed to search for showers of
zenith angle already above $60^0$; but it means that only the muon poor
fraction of the HAS produced by a hypothetical neutrino flux has to be compared
with this bound. Only the resonant channel decaying into an electron and an
electron-antineutrino produces a purely electromagnetic shower directly in the
interaction. Electron-neutrino charged current interactions in which the
hadronic component carries less than the $20\%$ of the initial energy
$E_{\nu}(y<0.2)$ produce muon poor showers according to AKENO requirements.
Since the average $<y>$ at high energies approaches 0.2\cite{Reno} a good
fraction of the electron neutrino showers are muon poor. Both these
contributions are calculated as described in the previous section. For the SCS
model we compare the total rate of HAS, to that from electron neutrinos and
antineutrinos only, separating the DIS and the resonant contributions in fig.
(\ref{contributions}). We also plot in the same figure the results for muon
poor horizontal showers as described above separating the resonant and DIS
channels as well. Of all the electron neutrino HAS, about $50\%$ are muon poor
at high energies.

Any channel for producing muons can also give pure electromagnetic showers
through secondary muon bremsstrahlung or electron positron production. In spite
of the suppression due to a double process its contribution can be relevant for
muon poor showers at large zenith angles. We consider muons produced in charged
current muon-neutrino interactions and those produced in the resonant
interaction decaying into a muon and a muon antineutrino, these muons carry
most of the energy of the interaction. Muons produced in hadronic showers or
tau decays have lower energy and can be neglected in comparison to other
uncertainties.  

For incoming muon neutrinos, we have firstly calculated the muon flux produced
by the charged current interaction of horizontal muon neutrino fluxes through
the 36000 $g~cm^{-2}$ slant depth.  This flux will give rise to purely
electromagnetic horizontal showers due to hard processes of energy loss such as
bremsstrahlung and electron positron production. These showers will generally
develop well after the first shower initiated by the nucleon debris  is
absorbed. The calculation of this channel uses eq.~(\ref{eq:shower}) with the
secondary muon flux, the muon cross section and $\bar y = y$. We consider the
dominant contribution, bremsstrahlung, and we use the cross section given by
\cite{Petrukhin}. The shower rate calculated in this way at $90^0$ represents
approximately $50\%$ of the "direct" electron neutrino showers with a cut in
$y<0.2$. Secondary bremsstrahlung is the most important channel for muon poor
showers induced by muon neutrinos and has a strong dependence on zenith angle.
The result of secondary muon bremsstrahlung HAS in the SCS model, averaged over
the zenith angles from $60^0$ to $90^0$, is also compared to the other channels
in fig.(\ref{contributions}). However the contribution represents less than a
$10\%$ effect on the total number of muon poor showers because of this
averaging.

\begin{figure}
\epsfxsize=10cm
\begin{center}
\mbox{\epsfig{file=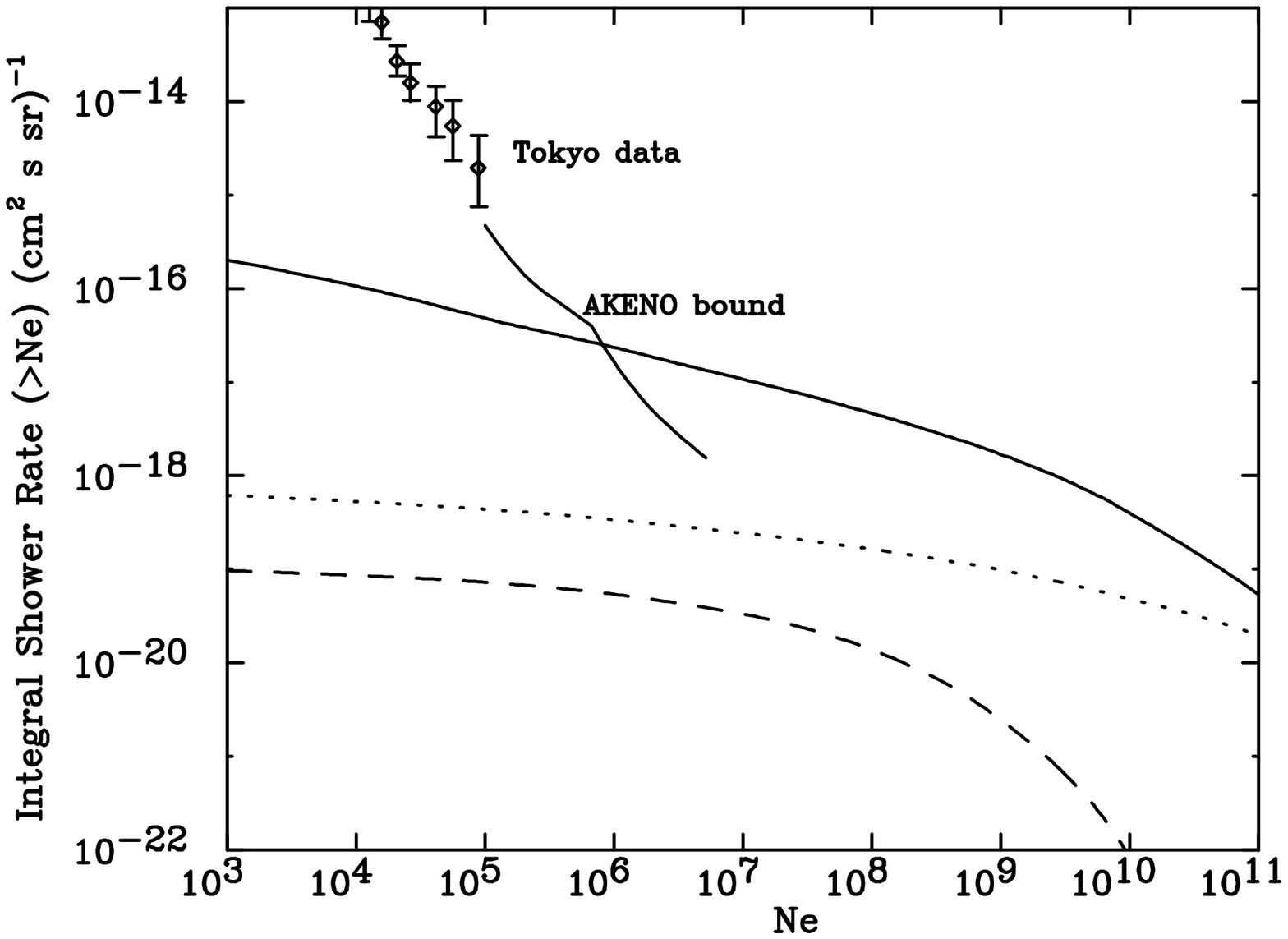}}
\end{center}
\caption{Muon poor integral shower rate for the three models.  TD $p=0$ (solid
line), $p=0.5$ (dotted) and AGN (dashed).} 
\label{showers2}
\end{figure}

In fig.(\ref{showers2}) we display the results for muon poor showers in the
considered models together with the AKENO bound where it is shown how the model
with $p=0$, corresponding to superconducting cosmic strings, is ruled out by an
ample margin. The figure also illustrates how the model with different time
evolution ($p=0.5$) is not far from being tested with current experiments
assuming improvements in statistics. Clearly searching for HAS of any type
should increase the expected signal, but care has to be taken to insure HAS are
not due to ordinary cosmic rays. We have include in fig.(\ref{showers}) a rough
theoretical estimate of the $90\%$ confidence upper limit for no observations
during a year, assuming no ordinary cosmic ray showers penetrate at angles
above $75^0$, for ideal air shower arrays of three areas $\sim 10^5~m^2$, $\sim
10^2~km^2$ and $\sim 10^4~km^2$. These are typical order of magnitude sizes of
existing and planned arrays.

{\sl Limits.} 
The information provided by Fr\`ejus, HAS and Fly's Eye, ruling  out the same
types of models, is complementary in the sense that they relate to very
different parts of the neutrino spectrum and because HAS apply to electron
neutrinos as well. It is interesting to compare the limit derived from the
AKENO bound with other limits for high energy neutrino fluxes. At lower
energies ($\sim$ 3 TeV), the Fr\`ejus limit \cite{Frejus} rules out the TD
prediction with $p=0$. The Fly's Eye limit \cite{Baltrusaitis}, also rules out
this prediction at much higher energies $\sim$ 10$^5$ TeV. The limit from AKENO 
lies in between, in the PeV region. The EAS-TOP collaboration \cite{EASTOP} has
also published a less constraining limit on high energy neutrino fluxes from
horizontal showers also near the PeV region. In fig.(\ref{fluxes}) we show the
flux predictions together with these limits. This figure, however, must be
interpreted with care as these limits can be very flux-dependent. We illustrate
this by plotting the limits corresponding to constant spectral index fluxes
with different indices ($\gamma=1.1,2,3$). In this respect it would be useful
that all experiments measuring Air Showers such as EAS-TOP, CYGNUS and Fly's
Eye (measuring upcoming showers), gave their results as air shower rates like
AKENO as this would simplify the comparisons between experiments and would be a
step forward to the establishment of these results.

\begin{figure}
\epsfxsize=10cm
\begin{center}
\mbox{\epsfig{file=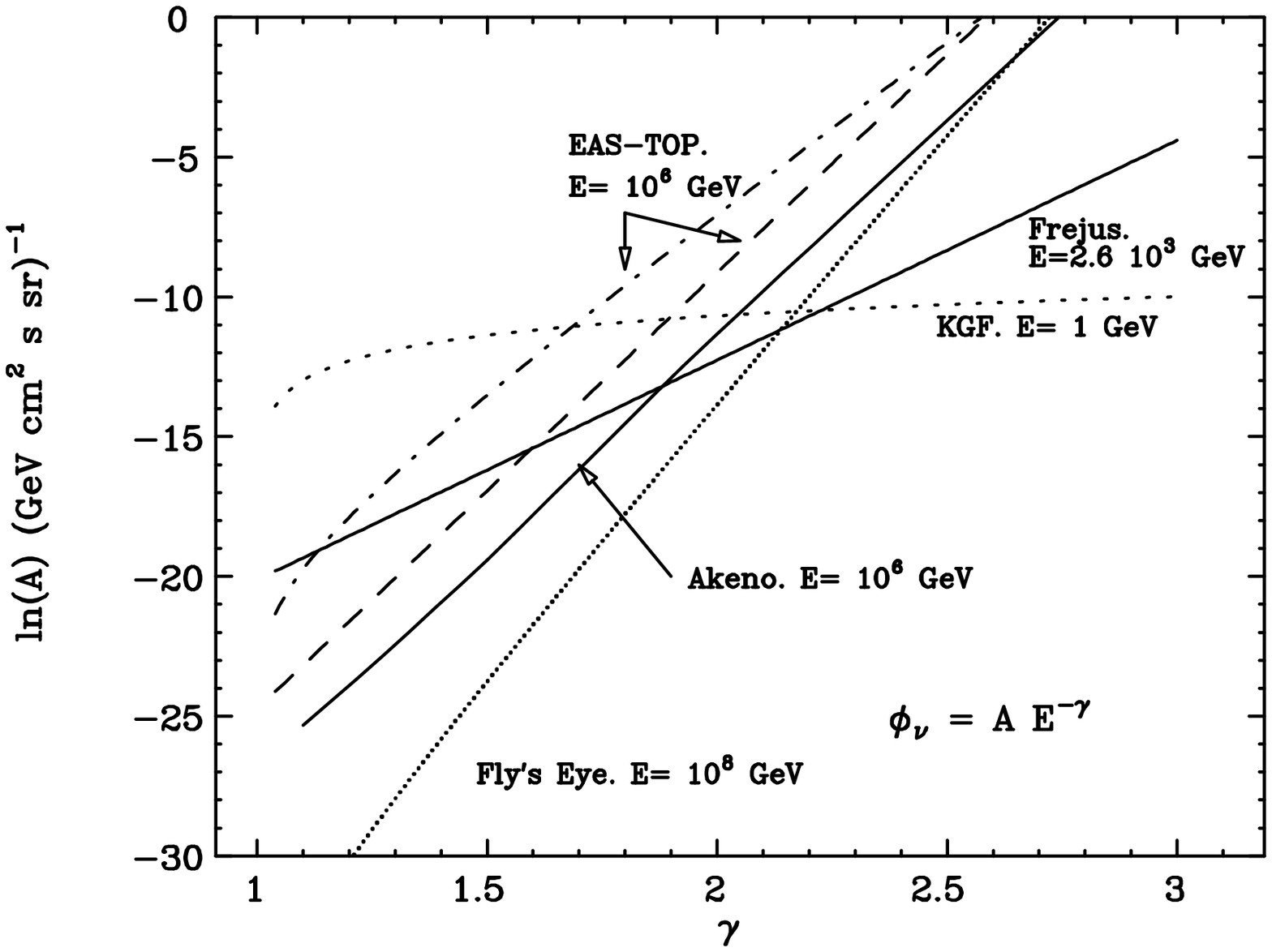}}
\end{center}
\caption{Limits on the $A - \gamma$ space (see text) given from different 
experiments, as marked in the figure. The energy attached to each line gives an 
indication of the energy involved in the corresponding experiment. The dotted 
line (marked as KGF) corresponds to the  point source limit (Markarian 421)
from the Kolar Gold Field experiment.}
\label{parlimit}
\end{figure}

To further illustrate this point we have considered a simple neutrino flux with
constant spectral index ($\gamma$) from the GeV to the EeV region $\phi(E)=A
E^{-\gamma}$.  For each of the above limits there is a curve in the $\gamma-A$
parameter space plane. This is shown in Fig.~(\ref{parlimit}). There are
several interesting things to point out from this figure. The best limit
corresponds to Fly's Eye for small spectral indices (as is the case for the TD
models), and to deep underground experiments such as Fr\`ejus if the index is
large. Besides, for $\gamma \sim $ 2 all experiments have similar performances
within an order of magnitude.

HAS (within which we include Fly's Eye results) are a tool for constraining TD
scenarios via the neutrino-induced showers produced along with the UHE cosmic
rays and they rule out superconducting cosmic string models normalized to the 
highest energy cosmic ray spectrum. The prospects of running the largest
existing air shower arrays using horizontal triggers, in order to get not only
the electromagnetic shower rate bounds, but the rate of all Horizontal Air
Showers, should further constrain the present TD models. Clearly the
possibility of searching for HAS with future air shower arrays such as the
Pierre Auger Project, with an area about 60 times larger than AKENO, if viable,
would undoubtfully provide a much stronger constraint on these models.

\centerline{\bf Acknowledgements}

We thank Jaime Alvarez Mu\~niz, Francis Halzen, Karl Mannheim, Gonzalo Parente
and G\"unter Sigl for helpful discussions. This work was partially supported by
CICYT under contract AEN96-1773, by the Xunta de Galicia under contract XUGA
20604A96 and by the EC (R.A.~V.), under contract ERBCHBICT941658.


\end{document}